\documentclass[11pt]{article}

\usepackage[utf8]{inputenc}
\usepackage[T1]{fontenc}
\usepackage{lmodern}
\usepackage[margin=1in]{geometry}
\usepackage{amsmath,amssymb}
\usepackage{graphicx}
\usepackage{booktabs}
\usepackage{tabularx}
\usepackage{array}
\usepackage{enumitem}
\usepackage[numbers,sort&compress]{natbib}
\usepackage{xcolor}
\usepackage{titlesec}

\definecolor{ResearchInk}{HTML}{20303A}
\definecolor{ResearchCopper}{HTML}{B86B4B}
\definecolor{ResearchSand}{HTML}{E8DED0}
\definecolor{LabtuneTeal}{HTML}{4E827C}
\definecolor{LabtuneBlue}{HTML}{607F98}
\definecolor{LabtuneMist}{HTML}{E4ECE9}
\definecolor{LabtuneSky}{HTML}{E6EDF2}
\definecolor{PaperGray}{HTML}{F3F1ED}

\titleformat{\section}
  {\large\bfseries\color{ResearchInk}}
  {\thesection}{0.65em}{}
  [\vspace{0.15em}{\color{ResearchCopper!55}\titlerule[0.45pt]}]
\titleformat{\subsection}
  {\normalsize\bfseries\color{LabtuneTeal}}
  {\thesubsection}{0.6em}{}
\titleformat{\paragraph}[runin]
  {\normalsize\bfseries\color{ResearchInk}}
  {}{0pt}{}[.]

\usepackage{tikz}
\usetikzlibrary{positioning,arrows.meta,fit,calc}
\usepackage{microtype}
\usepackage{hyperref}
\hypersetup{
  colorlinks=true,
  linkcolor=LabtuneBlue,
  citecolor=LabtuneTeal,
  urlcolor=ResearchCopper
}
\usepackage[capitalise,noabbrev]{cleveref}

\setlist{itemsep=2pt,topsep=4pt}
\newcolumntype{L}{>{\raggedright\arraybackslash}X}
\newcommand{\code}[1]{\texttt{#1}}

\title{\textcolor{ResearchInk}{\textbf{Labtune: Treating Lab Organization and Information Sharing\\as Experimental Variables in LLM-Agent Discovery}}}
\author{Goran Mitevski\\[2pt]
  \small \href{mailto:rolevski.mk@icloud.com}{\texttt{rolevski.mk@icloud.com}}}
\date{November 2026}

\begin{document}
\color{ResearchInk!96}
\maketitle

\begin{abstract}
De-siloing can create groupthink; silos can prevent knowledge sharing and reduce research
 efficiency. The same tension appears in research systems built with large language model
 (LLM) agents, but the organization of those systems is usually fixed inside the architecture.
 A pipeline, a debate, a principal investigator with specialists, or a densely communicating
 team is selected by the system designer before the research starts. I propose treating that
 choice as part of the experiment.

I introduce \emph{Labtune}, a framework and open reference implementation for running bounded
 research workflows with a user-selected laboratory organization, communication policy, model
 configuration and budget. A user can choose among six editable designs---multidisciplinary
 silos, transdisciplinary integration, complementary co-leaders, a hub-and-spoke matrix, a
 flat/holacratic co-lab, or a single scientist---and independently select how much information
 crosses team boundaries: none, low, intermediate or high detail, with an optional channel of
 conditional abstractions called \emph{wisdom}. Sharing is enforced by the controller at phase
 barriers rather than through unrestricted agent chat. Delegation does not create new budget or
 permissions. Development evidence is separated from protected confirmation, whose results do
 not return to the acting agents. An optional multi-lab layer connects laboratories with the
 same or different designs through similarly filtered exchanges, conferences and symposia.

This is an idea and system-description paper, not a benchmark result. The contribution is the
 framing of laboratory organization and information sharing as controlled variables, together
 with a concrete configuration space, an executable reference implementation, and hypotheses
 that can be tested under matched budgets. I do not claim that one lab design or one level of
 sharing is best. The point is to make those differences measurable.
\end{abstract}

\section{Introduction}
\label{sec:intro}

A research group can fail in two opposite directions. Too little exchange leaves useful
 knowledge trapped inside silos and allows teams to repeat the same mistakes. Too much exchange
 can collapse different lines of inquiry into one shared view before the problem has been
 sufficiently explored. Organizational research has described related tensions between
 exploration and exploitation \citep{march1991exploration}, between fast convergence and search
 diversity \citep{lazer2007network,mason2008propagation,zollman2010transient}, and between
 coordination and conformity \citep{janis1972groupthink}. Groups also tend to under-share
 information held by only one member \citep{stasser1985hidden}, while diversity of approaches
 can matter as much as individual ability \citep{hong2004diversity}. The useful question is
 therefore not whether silos or integration are good. It is \emph{what should be shared, with
 whom, and at what point in the search}.

LLM-agent research systems already make these choices. Some use a fixed sequence of stages;
 others generate and debate hypotheses, place a principal-investigator agent above specialists,
 or let a team discuss its way toward a result
\citep{lu2024aiscientist,schmidgall2025agentlab,gottweis2025coscientist,swanson2025virtuallab,boiko2023coscientist}.
The organization is normally treated as implementation detail rather than as a variable. This
is especially important when nominally different roles are instantiated by the same underlying
model. Different personas can then share similar priors, blind spots and stylistic tendencies.
Generative models can improve individual output while reducing collective diversity
\citep{doshi2024creativity}; recent work also reports cases in which multi-agent teams hold back
an expert member \citep{pappu2026experts} and fail through recurring coordination errors
\citep{cemri2025fail}. More communication is therefore not automatically more useful
collaboration.

The systems problem I address is simple: \textbf{if organization and information flow can alter
what an artificial research team explores, they should be recorded and varied with the same
care as the model, data and budget}. Labtune is a concrete attempt to make that possible. It has
five commitments:
\begin{itemize}
  \item \textbf{Selectable organization.} Six lab designs are supplied as editable Markdown
  files with machine-readable front matter describing personas, phases, governance,
  collaboration forms and delegation defaults (\cref{sec:designs}).
  \item \textbf{Sharing as a controlled policy.} Horizontal exchange passes through a
  controller-owned filter with four detail levels and an independent wisdom channel. Views are
  published at barriers and constrained by a read budget (\cref{sec:sharing}).
  \item \textbf{Conservation of budget and permissions.} A child agent consumes the parent's
  allocation and cannot widen the parent's access (\cref{sec:agents}).
  \item \textbf{Evidence separation.} Agents work on development evidence. A fixed candidate is
  later evaluated on protected confirmation data that does not return to the agents. The user
  selects frequentist or Bayesian analysis, with a reviewed option for model-assisted prior
  elicitation (\cref{sec:evidence}).
  \item \textbf{No automatic redesign.} Labtune does not search for a winning organization or
  silently change communication settings. The user chooses; the system records
  (\cref{sec:principles}).
\end{itemize}

\paragraph{Scope of this paper} This is an exploratory preprint and system description. I
contribute (i) the framing of lab design and information sharing as controlled variables for
LLM-agent research, (ii) design principles and a formal configuration space, (iii) an open
reference implementation \citep{labtune2026code} with an executable synthetic research task,
and (iv) falsifiable hypotheses and a protocol for testing them (\cref{sec:agenda}). I report no
comparative result here. If one design appears better in a future study, that result should be
understood as conditional on the mission, models, budgets, evidence and communication policy
used in that study.

\section{Background and related work}
\label{sec:related}

\paragraph{Organization, networks and exploration} March's exploration--exploitation
 distinction \citep{march1991exploration} provides the broad organizational background. In
 networked search, fast information flow can spread good solutions quickly but can also lock a
 group into a mediocre region of the search space \citep{lazer2007network}; human experiments
 show related effects \citep{mason2008propagation}. In social epistemology, transient diversity
 can be useful because a community has time to maintain competing beliefs before convergence
 \citep{zollman2010transient}. Scientific team structure is also associated with the character
 of research output \citep{wu2019large}. Labtune borrows vocabulary from multidisciplinary,
 interdisciplinary and transdisciplinary work \citep{rosenfield1992continuum}, and from
 role-based self-management \citep{robertson2015holacracy}, but does not assume that human
 organizational results transfer directly to LLM agents. The framework is meant to make that
 transfer testable rather than implicit.

\paragraph{LLM agents for research} Existing systems already instantiate different laboratory
 forms. The AI Scientist automates a path from idea to paper \citep{lu2024aiscientist}; Agent
 Laboratory structures literature review, experimentation and writing with human feedback
 \citep{schmidgall2025agentlab}; the AI co-scientist uses a generate--debate--evolve process
 \citep{gottweis2025coscientist}; the Virtual Lab places a principal-investigator agent above
 specialists who meet and exchange results \citep{swanson2025virtuallab}; and Coscientist links
 planning to laboratory automation \citep{boiko2023coscientist}. Human evaluation also suggests
 that LLM-generated research ideas can be judged more novel but less feasible than expert ideas
 \citep{si2024ideas}. These systems answer the organization question differently. Labtune's
 contribution is to expose that answer as a setting that can be held fixed, changed and compared.

\paragraph{Multi-agent LLM coordination} Role-playing and software-team frameworks
 \citep{li2023camel,hong2023metagpt,qian2023chatdev,wu2023autogen}, together with multi-agent
 debate \citep{du2023debate,liang2023divergent}, show that structured interaction can improve
 some outputs. Work on collaboration strategies \citep{zhang2023socialpsych} and communication
 topology \citep{qian2024macnet} further suggests that the pattern of interaction matters.
 Failure analyses \citep{cemri2025fail} and evidence that teams can hold experts back
 \citep{pappu2026experts} motivate a narrower question: not only \emph{who communicates}, but
 \emph{what information is permitted to cross the boundary}.

\paragraph{Automated design of agent organizations} Other work searches automatically over
 agent architectures or communication graphs \citep{hu2024adas,zhuge2024gptswarm}, and IMACS
 routes tasks to collaboration protocols with a contextual bandit \citep{chen2026imacs}. Labtune
 deliberately takes the opposite position during a research run. If the organization is adapted
 against development signals, organization and optimization become difficult to separate. I
 therefore keep the organizational configuration user-selected and stable, with changes made
 between recorded runs. The approaches are still compatible: an automated designer can propose
 configurations, which can then be evaluated as fixed Labtune conditions under matched budgets.

\section{Design principles}
\label{sec:principles}

The implementation follows seven rules. They are intentionally conservative because the goal is
not to make the agents appear autonomous; it is to make differences between research conditions
observable.

\begin{enumerate}[label=\textbf{P\arabic*.},leftmargin=*]
  \item \textbf{The user chooses the organization.} Lab design, persona counts, delegation,
  sharing, models, budgets and statistical approach are explicit. Presets may suggest defaults,
  but they never silently raise a user-selected sharing level or budget. Precedence is: mission
  and runtime limits $\rightarrow$ explicit user choices $\rightarrow$ preset suggestions
  $\rightarrow$ base defaults.
  \item \textbf{Sharing is a policy, not a side effect.} Information crosses team boundaries
  only through controller-published, recipient-specific views. A permitted link does not grant
  access to the linked artifact, and a coordinator with broader access cannot relay richer
  content to a restricted recipient.
  \item \textbf{Delegation conserves resources and permissions.} Spawning a child creates no
  additional budget, vote or access.
  \item \textbf{Evidence outranks assertion.} Measured claims require evaluator receipts. A
  model's confidence is a judgment, not an experimental probability; p-values, intervals and
  posteriors come only from identified analysis tools.
  \item \textbf{Confirmation is protected.} The candidate is fixed before confirmation.
  Confirmation uses disjoint data and does not feed results back to acting agents or later
  sequences.
  \item \textbf{Exposure history is part of the result.} Configuration identities include model
  and endpoint bindings. A later change in sharing policy cannot erase information already read,
  so mixed-exposure runs remain labelled as such.
  \item \textbf{The system does not improve its own organization.} There is no hidden redesign
  loop, prompt evolution or winner promotion. The user runs and interprets the comparison.
\end{enumerate}

\section{The Labtune model}
\label{sec:model}

\begin{figure}[t]
\centering
\begin{tikzpicture}[
  font=\scriptsize,
  stp/.style={draw, rounded corners=2pt, align=center, text width=2.15cm, minimum height=1.6cm, inner sep=3pt},
  sup/.style={draw, rounded corners=2pt, align=center, text width=2.15cm, minimum height=1.5cm, inner sep=3pt, fill=ResearchInk!4},
  arr/.style={-{Latex[length=1.6mm]}, thick},
  up/.style={-{Latex[length=1.6mm]}}
]
\node[stp, fill=ResearchInk!5] (s1) {\textbf{Mission + lab}\\[1pt] choose 1 of 6 editable designs};
\node[stp, fill=LabtuneTeal!10, right=0.25cm of s1] (s2) {\textbf{Orchestrate}\\[1pt] dependency waves, global budgets};
\node[stp, fill=LabtuneBlue!11, right=0.25cm of s2] (s3) {\textbf{Researchers}\\[1pt] one scientist or bounded teams};
\node[stp, fill=LabtuneTeal!10, right=0.25cm of s3] (s4) {\textbf{Execute + measure}\\[1pt] qualified sandbox and evaluator};
\node[stp, fill=LabtuneTeal!10, right=0.25cm of s4] (s5) {\textbf{Confirm}\\[1pt] reserved data; no agent feedback};
\node[stp, fill=ResearchInk!5, right=0.25cm of s5] (s6) {\textbf{Candidate + receipts}\\[1pt] measured trade-offs and limits};
\foreach \a/\b in {s1/s2,s2/s3,s3/s4,s4/s5,s5/s6} \draw[arr] (\a) -- (\b);
\node[sup, below=1.15cm of s1] (b1) {\textbf{User}\\ selects settings; edits between runs};
\node[sup, below=1.15cm of s2, fill=ResearchCopper!10] (b2) {\textbf{Configuration identity}\\ no automatic redesign};
\node[sup, below=1.15cm of s3, fill=ResearchCopper!10] (b3) {\textbf{Horizontal views}\\ none / low / int.\ / high $\pm$ wisdom};
\node[sup, below=1.15cm of s4, fill=ResearchSand!42] (b4) {\textbf{Optional decisions}\\ LLM-only / shadow / active};
\node[sup, below=1.15cm of s5, fill=LabtuneTeal!7] (b5) {\textbf{Statistics}\\ frequentist or Bayesian (user)};
\node[sup, below=1.15cm of s6, dashed, fill=white] (b6) {\textbf{Asset lifecycle}\\ integration, registry (planned)};
\foreach \a/\b in {b1/s1,b2/s2,b3/s3,b4/s4,b5/s5} \draw[up] (\a) -- (\b);
\draw[up, dashed] (s6) -- (b6);
\draw[arr, ResearchInk!65] (s4.north) -- ++(0,0.45) -| node[pos=0.25, above, font=\scriptsize\itshape]{development evidence $\rightarrow$ candidate refinement} (s3.north);
\end{tikzpicture}
\caption{A Labtune lab from mission to measured output. Top row: the research flow. Bottom
row: the records and user choices that govern each step. Development evidence refines
candidates but never changes the user's organization; confirmation results never return to
agents. The general asset lifecycle (dashed) is planned.}
\label{fig:overview}
\end{figure}
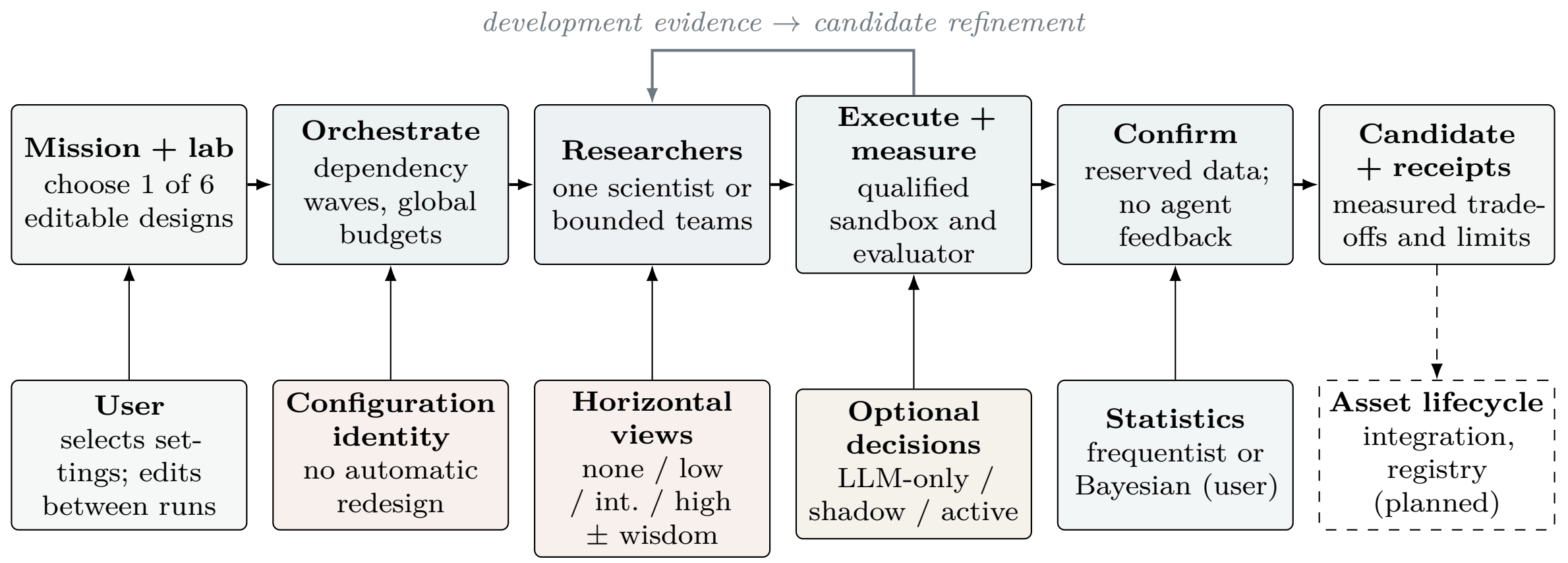

\Cref{fig:overview} shows the basic loop. The user supplies a mission and chooses a lab
configuration. The controller turns that configuration into a dependency graph, schedules work
within global limits, and gives configured agents access only to their permitted tools and
views. Qualified domain adapters execute and measure experiments. A reserved confirmation stage
then evaluates the fixed candidate.

\subsection{Configuration space}
\label{sec:config}

A lab configuration is a tuple
\begin{equation}
  \mathcal{L} = (D,\; \mathcal{P},\; \Delta,\; \pi,\; \mathcal{M},\; \mathcal{B},\; \sigma),
\end{equation}
where $D$ is a lab design (\cref{sec:designs}); $\mathcal{P}$ the persona-to-agent assignment;
$\Delta = (d_{\max}, c_{\max}, a_{\max})$ the delegation limits (maximum depth, children per
agent, active descendants); $\pi = (\ell, w)$ the horizontal sharing policy with detail level
$\ell \in \{\textsf{none},\textsf{low},\textsf{intermediate},\textsf{high}\}$ and wisdom flag
$w \in \{0,1\}$ (\cref{sec:sharing}); $\mathcal{M}$ the generation model and optional decision
endpoints; $\mathcal{B}$ the budgets (requests, wall time, active agents, experiments); and
$\sigma \in \{\textsf{frequentist},\textsf{bayesian}\}$ the statistical approach. The
configuration is loaded from strict YAML, validated against JSON schemas, and hashed into a
configuration identity that is stored with every plan and run. Planning, validation and
visualization make no model calls, so the configuration space can be inspected for free.

\subsection{Lab designs}
\label{sec:designs}

A design is an editable Markdown file with YAML front matter. It declares governance,
personas, methodological approaches, phases, collaboration artifacts, recommended sharing,
meeting limits and delegation defaults. The prose body describes priorities, expected outputs,
ways of handling disagreement and failure modes to watch. \Cref{tab:designs} lists the six
presets. A user can edit them or add a new design from a template, but a new label is not enough:
the design must change how work is assigned, how evidence moves, how disagreement is treated, or
how partial results are integrated.

\begin{table}[t]
\centering
\small
\caption{The six lab-design presets. Suggested sharing is a default the user may override; it
is not a claim of optimality.}
\label{tab:designs}
\begin{tabularx}{\linewidth}{@{}l L L l@{}}
\toprule
Design & Organization and personas & Phases & Suggested $\pi$ \\
\midrule
A \;Multidisciplinary silos & Formal, empirical and implementation specialists pursue separate tracks & Independent framing, parallel episodes, independent verification, optional handoff & low + wisdom \\
B \;Transdisciplinary integration & Structural modeler, experimentalist, systems designer, plus a method integrator & Independent framing, translation workshop, parallel tests, framework revision & high + wisdom \\
C \;Complementary co-leaders & A mechanism lead and a measurement lead, each with a team & Independent positions, parallel episodes, paired challenge, discriminating test & intermediate + wisdom \\
D \;Hub-and-spoke matrix & A research lead coordinating creative, technical and measurement tracks & Track charters, parallel episodes, hub review, scoped handoffs & intermediate + wisdom \\
E \;Flat / holacratic co-lab & Exploratory, builder and critical peers; rotating facilitator & Independent proposals, roundtable, task adoption, evidence review & intermediate + wisdom \\
F \;Single scientist & One scientist with low, medium or persistent effort & Investigate, then review and refine & none \\
\bottomrule
\end{tabularx}
\end{table}

Two distinctions are important. The \emph{code controller} owns queues, budgets, permissions,
evidence integrity and publication; a \emph{research leader} is only an LLM role that proposes
direction. A hub lead or flat-lab facilitator cannot rewrite evidence or create resources.
Second, a design never overrides a stricter sharing choice. If a transdisciplinary lab is run
with \textsf{none}, it keeps the independent framings and postpones synthesis. The resulting run
is recorded as a restricted variant and should not be described as if full integration occurred.

\subsection{Agents, delegation and scheduling}
\label{sec:agents}

Each design instantiates \emph{prime} agents at depth 0, with each prime assigned a distinct
approach to the mission. A prime can delegate bounded subtasks to children at depth 1, and deeper
only when $d_{\max} \geq 2$. Let $T(p)$ be the team rooted at prime $p$ and $c(a)$ the resources
consumed by agent $a$. The controller enforces
\begin{equation}
  \sum_{a \in T(p)} c(a) \;\le\; B_p, \qquad \sum_{p} \sum_{a \in T(p)} c(a) \;\le\; B_{\text{lab}},
\end{equation}
together with separate caps on concurrently active agents, LLM requests and experiment jobs.
Topology does not create capacity. Excess work queues, and a waiting parent yields capacity so
its child can run. Children inherit the parent's sharing permissions and may only narrow them.
Their findings return to the parent, which can propose them for publication. Work is executed in
dependency-ready waves, actor contexts remain separate, and peer publication occurs only after a
phase barrier.

\paragraph{Decisions inside an episode.}
An agent can pause an episode, ask a bounded question through \code{decision.assess}, and then
continue. The user chooses \emph{LLM-only}, where the generation model answers; \emph{shadow},
where a separate endpoint records an opinion that is not returned to the acting context; or
\emph{selected-active}, where a valid specialist answer is returned and an unavailable or
invalid answer may fall back to the generation model. Every mode uses the same remaining
budget, sees only the caller's permitted context, and records provider identity, resolved model,
tokens and timing. A decision remains a decision; it is never promoted to empirical evidence.

\paragraph{Single scientist.}
Design F contains one prime and no children. The user selects \textsf{low}, \textsf{medium}
or \textsf{persistent} effort: in fixed-pass mode, 1, 3 or 6 passes per sequence, together with
a sequence count $S$. Later passes must do more than restate the first answer; they audit prior
work, address a concrete weakness or test an alternative. In persistent budget mode, sequence
$s$ receives an equal share of the remaining allowance, $R_s/(S-s+1)$. Retries and fallbacks do
not create new budget. Context continues by default, with an explicit fresh-context option.
This is the single-agent baseline for multi-agent comparisons.

\section{Horizontal information sharing}
\label{sec:sharing}

\subsection{Detail levels and the wisdom channel}

Each prime team writes to a private structured outbox: a summary, typed findings, an optional
candidate, and optional wisdom entries. Dead ends can carry evidence references. At a barrier,
the controller applies a filter $V_{\pi}$ to peer outputs and publishes an immutable, versioned
Markdown view for each recipient. If $o_{-i}(t)$ denotes peers' outputs at barrier $t$, the
context of team $i$ evolves as
\begin{equation}
  C_i(t+1) \;=\; C_i(t) \,\cup\, \operatorname{trunc}_W\!\big(V_{\pi}(o_{-i}(t))\big),
\end{equation}
where $\operatorname{trunc}_W$ limits the combined detail and wisdom view to a read budget of
$W$ words. Agents never write into a common research file. The event store remains authoritative;
the Markdown is the exact recipient view and is retained with hashes for replay. \Cref{tab:sharing}
defines the levels.

\begin{table}[t]
\centering
\small
\caption{Horizontal sharing levels. Detail and wisdom are independent settings, giving eight
useful configurations: the five named modes plus low, intermediate or high with wisdom.}
\label{tab:sharing}
\begin{tabularx}{\linewidth}{@{}l L L@{}}
\toprule
Mode & Admitted content & Intended trade-off \\
\midrule
\textsf{none} & No peer research content; operational scheduling only & Baseline that preserves independent search \\
\textsf{low} & Notable dead ends only, each backed by a valid development receipt; no positive results, rankings or recipes & Avoid repeating costly failures without steering toward a solution \\
\textsf{intermediate} & Compact findings, broad method descriptions, constraints and trade-offs; no exact code or tuning recipes & Share useful direction while limiting copying \\
\textsf{high} & Experiment summaries, measurements, methods, parameters, permitted artifact references & Deliberate reuse and recombination \\
\textsf{wisdom\_only} & Conditional abstractions only (lesson, conditions, limits) & General lessons with reduced solution steering \\
\bottomrule
\end{tabularx}
\end{table}

The \textbf{wisdom} channel is an attempt to separate a lesson from the recipe that produced
it. A wisdom entry records a principle suggested by evidence, the conditions under which it
applies, its limits and exceptions, an evidence status (tentative, repeated within a scope,
contradicted, superseded), and evidence identifiers. The recipient sees only the abstraction.
Prime identity, ranking, exact parameters, implementation recipes and suggested next actions are
withheld, while the controller retains full attribution. For example, a lab may learn that
comparing interventions after the same number of iterations can imply unequal resource use
without revealing the implementation that exposed the problem. Whether this transfers useful
structure while preserving independent search is a hypothesis, not a property guaranteed by
the format (\cref{sec:agenda}).

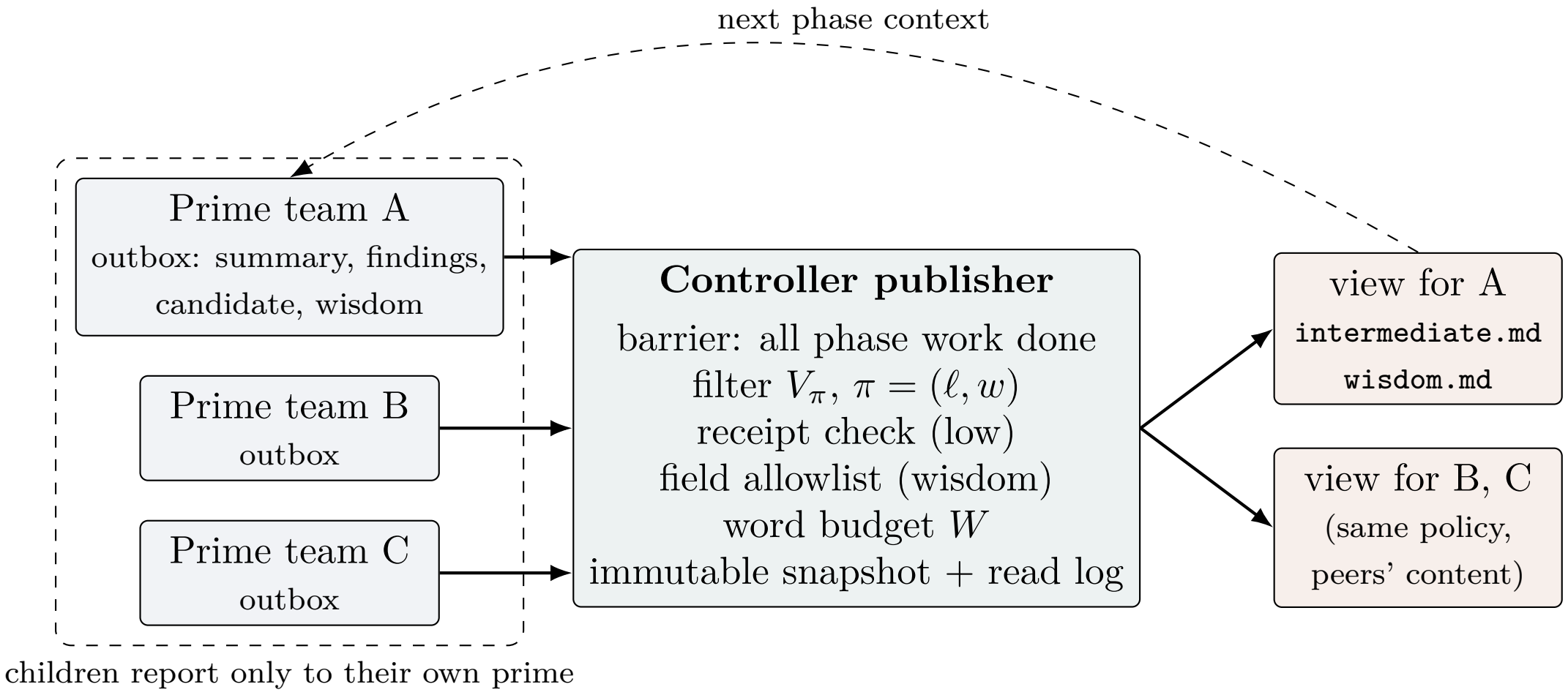
\begin{figure}[t]
\centering
\begin{tikzpicture}[
  font=\small,
  box/.style={draw, rounded corners=2pt, align=center, inner sep=4pt},
  team/.style={box, fill=LabtuneBlue!9, minimum width=2.7cm},
  ctrl/.style={box, fill=LabtuneTeal!10, minimum width=4.3cm, minimum height=2.5cm},
  view/.style={box, fill=ResearchCopper!11, minimum width=2.6cm},
  arr/.style={-{Latex[length=2mm]}, thick},
  darr/.style={-{Latex[length=2mm]}, dashed}
]
\node[team] (a) {Prime team A\\\scriptsize outbox: summary, findings,\\\scriptsize candidate, wisdom};
\node[team, below=0.35cm of a] (b) {Prime team B\\\scriptsize outbox};
\node[team, below=0.35cm of b] (c) {Prime team C\\\scriptsize outbox};
\node[ctrl, right=1.2cm of b] (p) {\textbf{Controller publisher}\\[3pt]
  barrier: all phase work done\\
  filter $V_{\pi}$, $\pi=(\ell,w)$\\
  receipt check (low)\\
  field allowlist (wisdom)\\
  word budget $W$\\
  immutable snapshot + read log};
\node[view, right=1.2cm of p, yshift=0.9cm] (v1) {view for A\\\scriptsize \code{intermediate.md}\\\scriptsize \code{wisdom.md}};
\node[view, right=1.2cm of p, yshift=-0.9cm] (v2) {view for B, C\\\scriptsize (same policy,\\\scriptsize peers' content)};
\draw[arr] (a.east) -- (p.west |- a.east);
\draw[arr] (b.east) -- (p.west);
\draw[arr] (c.east) -- (p.west |- c.east);
\draw[arr] (p.east) -- (v1.west);
\draw[arr] (p.east) -- (v2.west);
\node[draw, dashed, rounded corners, fit=(a)(b)(c), inner sep=5pt, label={[font=\scriptsize]below:children report only to their own prime}] {};
\draw[darr] (v1.north) to[out=150,in=30] node[above, font=\scriptsize]{next phase context} (a.north);
\end{tikzpicture}
\caption{Controller-mediated horizontal sharing. Teams write only to private outboxes; at a
barrier the publisher builds recipient views under the selected policy. In the reference
implementation, \textsf{low} admits only dead-end findings whose evidence references match
valid development receipts, and wisdom admits only the \emph{lesson}, \emph{conditions} and
\emph{limits} fields.}
\label{fig:sharing}
\end{figure}

\subsection{Meetings and other collaboration forms}

Designs B--E add bounded collaboration forms: bulletins, roundtables, workshops, proposal and
objection records, a versioned conceptual framework, paired challenges and handoffs. These forms
change \emph{how} permitted information is exchanged, not \emph{how much}. A meeting is a
scheduled task rather than an always-on chat. Participants first commit a private position,
then receive a size-limited packet allowed by the sharing policy. Each contributes one
observation, objection or costed proposal, and a facilitator records agreement, unresolved
disagreement and the next test. Round and word limits are explicit. Consensus without evidence
remains unsupported, and minority views stay in the record.

\subsection{Multi-lab studies}
\label{sec:multilab}

\begin{figure}[t]
\centering
\begin{tikzpicture}[
  font=\scriptsize,
  b/.style={draw, rounded corners=2pt, align=center, text width=2.3cm, minimum height=1.1cm, inner sep=3pt},
  arr/.style={-{Latex[length=1.6mm]}, thick}
]
\node[b, fill=ResearchInk!5] (u) {\textbf{User-selected collection}\\ 2--32 lab variants, $S$ sequences, meeting type};
\node[b, fill=LabtuneBlue!11, right=0.5cm of u, yshift=0.75cm] (l1) {\textbf{Lab 1}\\ design $D_1$; own context and evidence};
\node[b, fill=LabtuneBlue!11, right=0.5cm of u, yshift=-0.75cm] (l2) {\textbf{Lab 2..N}\\ same or different designs};
\node[b, fill=ResearchCopper!10, right=0.5cm of l1, yshift=-0.75cm] (x) {\textbf{Barrier + filtered exchange}\\ frozen snapshot; inter-lab $\pi$, word budget};
\node[b, fill=LabtuneBlue!8, right=0.4cm of x] (m) {\textbf{Conference / symposium}\\ presentations; 0--20 rounds; re-filtered};
\node[b, fill=LabtuneTeal!10, right=0.75cm of m] (c) {\textbf{Per-lab confirmation}\\ after all sequences and meetings};
\draw[arr] (u.east) -- (l1.west);
\draw[arr] (u.east) -- (l2.west);
\draw[arr] (l1.east) -- (x.west);
\draw[arr] (l2.east) -- (x.west);
\draw[arr] (x) -- (m);
\draw[arr] (m) -- node[pos=0.72, above, font=\scriptsize]{$s = S$} (c);
\draw[arr, dashed, ResearchInk!72] (m.south) -- ++(0,-1.05) -| node[pos=0.25, below, font=\scriptsize]{$s < S$: next sequence (continued or fresh context, plus permitted peer views)} (l2.south);
\node[draw, dotted, rounded corners, fit=(l1)(l2)(x)(m), inner sep=4pt, label={[font=\scriptsize]above:group and per-lab budgets cover research, meetings and fallbacks}] {};
\end{tikzpicture}
\caption{Multi-lab studies. The user selects a collection of lab variants, the inter-lab
sharing policy, a meeting type and a sequence count. All labs finish each research stage
before exchange; every stage reads a frozen snapshot of the previous one; and confirmation for
each lab waits until every sequence and meeting has ended.}
\label{fig:multilab}
\end{figure}
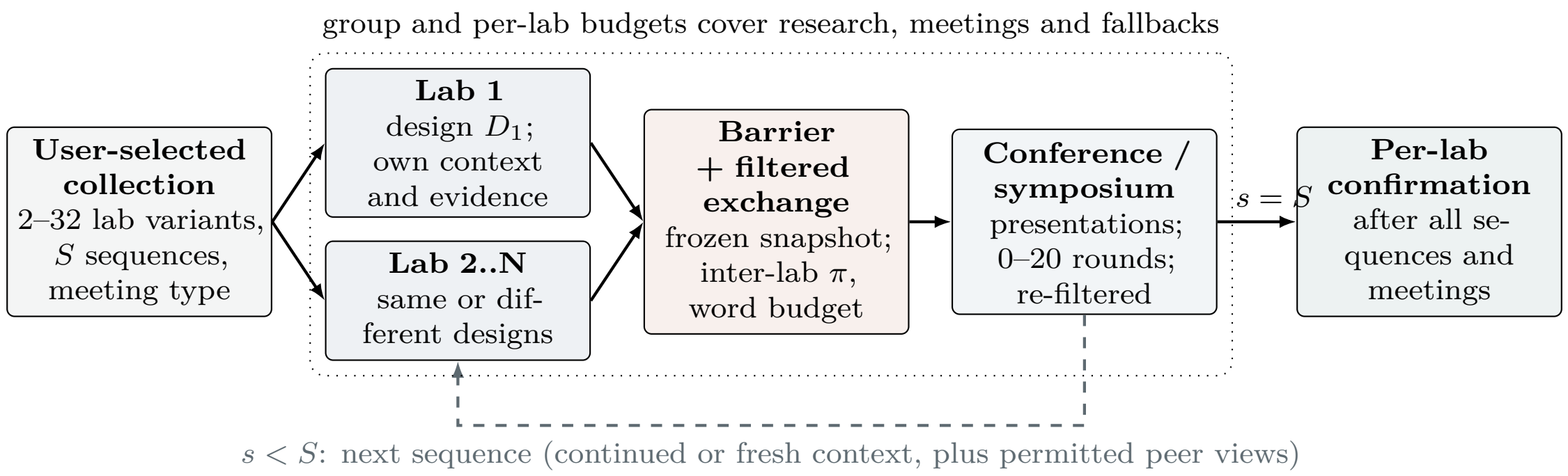

The same idea can be moved one level up. An optional layer composes 2--32 labs, using the same
or different designs, while preserving separate histories, evidence and confirmation
(\cref{fig:multilab}). Inter-lab sharing is a separate export permission applied at sequence and
meeting boundaries. Every lab completes a stage before exchange, and the next stage reads a
frozen snapshot, so completion order does not determine what was visible. Labs may hold a
\emph{conference}, with one presentation per lab, or a \emph{symposium}, with presentations and
1--20 bounded discussion rounds. Contributions are filtered again before another lab receives
them. With \textsf{none} and no wisdom, a meeting is operational only and makes no model calls.
Group and per-lab budgets cover research, meetings and fallbacks, and confirmation begins only
after all sequences and meetings finish. Different lab identifiers do not create independent
replications when the labs share a mission, model or private evaluation panel.

\section{Evidence, statistics and confirmation}
\label{sec:evidence}

\paragraph{Receipts and observable transitions.}
The useful unit of record is an observable transition: question $\rightarrow$ prediction and
proposed action $\rightarrow$ actual tool invocation and receipt $\rightarrow$ observation
$\rightarrow$ revised hypothesis or scoped claim. Each transition is linked to the actor, team,
configuration, input snapshot, resource use and artifact hashes. This keeps a proposal separate
from a test and a post-hoc explanation separate from an observation. Memory is layered into an
immutable evidence archive, operational playbooks and recipient-filtered horizontal views.
Wisdom is derived from retained evidence; it never replaces that evidence.

\paragraph{User-selected statistical family.}
Statistical family is independent of organization. The user chooses frequentist or Bayesian
analysis, while the domain adapter specifies the actual question, unit, comparator, effect,
replication, assumptions and stopping rule before the data are analyzed. No default significance
threshold or posterior decision criterion is imposed. Each receipt records whether statistics
were not required, not run, unsupported, completed or inconclusive. An unavailable analysis is
therefore not presented as a completed one or silently replaced by another family. This is
consistent with guidance that a p-value is not an effect size or the probability that a
hypothesis is true \citep{wasserstein2016asa}, and with the logic of declaring analyses before
seeing the outcome \citep{nosek2018prereg}.

\paragraph{Reviewed model-assisted priors.}
For Bayesian binomial analyses, a user-selected model can \emph{recommend} a prior but cannot
approve it. The user supplies at least three candidate Beta distributions with rationales and
separates evidence into three roles: prior evidence shown to the model, independent validation
counts used for prior predictive checks, and likelihood data reserved for the analysis and kept
out of the model context. A checker rejects known overlap through shared paths, identical hashes,
intersecting lineage identifiers or overlapping observations. Diagnostics report the prior
mean, concentration, a 95\% prior interval, predictive probability outside a declared plausible
range, and conservative two-sided predictive tails for validation counts, following the spirit
of prior predictive checking \citep{gabry2019visualization,gelman2013bda}. A named reviewer must
approve a passing candidate. Results show the selected posterior alongside all declared
alternatives and flag sensitivity when conclusions differ. Model confidence never determines
prior strength. The checks detect declared overlap, not unknown aliases or training-data
contamination.

\paragraph{Development and protected confirmation.}
Agents see development evidence only. The candidate is fixed before confirmation. The
controller then evaluates it on disjoint data, and those results do not return to the acting
agents or later sequences. Statistics computed after adaptive development are labelled
exploratory rather than treated as protected confirmation.

\section{Reference implementation}
\label{sec:impl}

The reference implementation, Labtune 0.5.0, is a Python ($\geq$3.11) package. Its core depends only on PyYAML and
\code{jsonschema} \citep{labtune2026code}. \Cref{tab:impl} separates implemented components from
planned interfaces. \code{labtune init} creates an independent workspace from packaged starters;
\code{prepare} validates and plans without credentials; and an explicit \code{run} begins a
budgeted live study. Generation uses a user-configured chat-completions endpoint, while optional
decision endpoints can use chat or typed JSON protocols. No provider, model or account is
bundled or preferred. Placeholder endpoints are valid for planning but rejected before a live
request, remote connections require HTTPS, and failed requests consume budget rather than being
retried invisibly.

\begin{table}[t]
\centering
\small
\caption{Implementation status of Labtune 0.5.0.}
\label{tab:impl}
\begin{tabularx}{\linewidth}{@{}l L L@{}}
\toprule
Layer & Implemented & Not implemented / limits \\
\midrule
Structural core & Six designs, schemas, strict YAML, phase DAGs, configuration identities & --- \\
Study runner & Explicit provider calls, bounded concurrency and delegation, JSON correction, filtered Markdown sharing, receipts & Live mid-run reconfiguration; durable scheduler recovery \\
Multi-lab & 2--32 labs, frozen exchanges, conferences, symposia, sequences, group budgets & No pooled quality score or winner \\
Statistics & Clopper--Pearson intervals; Beta--binomial posteriors; reviewed priors & Other likelihoods are adapter-specific \\
Domain tools & Registered adapters; sandboxed synthetic memory-policy task & Candidate execution only on macOS sandbox \\
Visualization & Local read-only setup visualizer and runtime observer; offline HTML snapshots & Shows metadata, not prompts or results \\
Asset lifecycle & Schemas and contracts & General integration and registry \\
\bottomrule
\end{tabularx}
\end{table}

\paragraph{Observability.}
Two local, read-only visualization agents show the resolved process graph and filtered runtime
metadata: which actor or request is active and which barrier has been reached. They make no model
calls and expose no prompts, generated code, private seeds or results. Live activity is kept
visually distinct from recorded or stale observations.

\paragraph{An executable research task: memory retention.}
The repository includes one synthetic task to exercise the complete path from authored source
to measured evidence. Agents are asked to write a causal retention policy for a streaming
nearest-neighbor anomaly detector under the same persistent-state budget as FIFO and reservoir
baselines. A useful policy should retain infrequent normal patterns, adapt to benign drift and
returning regimes, and remain sensitive to faults. Candidates implement
\code{select(bank, ages, new, t, capacity, aux)}, returning retained indices and up to 32
auxiliary numbers; with capacity 24 and dimension 16 the logical numeric state is at most
$24 \cdot (16+1) \cdot 8 + 32 \cdot 8 = 3{,}520$ bytes. A trusted supervisor holds labels,
private seeds and future inputs; submitted source executes only in a separate,
AST-constrained, OS-sandboxed interpreter with no file writes, network or process creation, and
with CPU, time and memory monitoring. Detection thresholds are calibrated on maxima of
separate event-free runs at rank $\lceil (n+1)(1-\alpha) \rceil$, which under exchangeability
gives a marginal bound $\alpha$ on the probability of any alarm in a benign finite-horizon run
\citep{vovk2005alrw}. Primary measures are run-level proportions (any false alarm per benign run,
fault detected within its interval, false alarm on benign recurrences) with exact
Clopper--Pearson 95\% intervals \citep{clopper1934}, plus delay, state bytes, process memory,
CPU and wall time; failed candidates remain visible. Calibration, development and confirmation
use disjoint private seed panels. A small smoke profile checks operation; a separate full
profile (200 calibration runs, 100 null and 100 fault runs, targets $\alpha \in \{0.10, 0.03,
0.01\}$, and 8, 16 or 40 normal modes, sometimes exceeding capacity) is intended for actual
comparisons. The smoke profile demonstrates that the loop executes. It is not evidence that the proposed
research method succeeds.

\section{Research agenda: hypotheses and protocol}
\label{sec:agenda}

Labtune is useful only if it makes organizational claims falsifiable. The following hypotheses
are therefore stated as possibilities, not expectations. Communication that helps one mission
may reduce performance on another.

\begin{enumerate}[label=\textbf{H\arabic*.},leftmargin=*]
  \item \textbf{Dead-end sharing.} Relative to \textsf{none}, \textsf{low} sharing reduces
  repeated costly failures without reducing the number of distinct mechanisms tested.
  \item \textbf{Wisdom transfer.} \textsf{wisdom\_only} transfers useful problem structure
  (improving verified outcomes per unit budget relative to \textsf{none}) while preserving
  more methodological diversity than \textsf{intermediate} or \textsf{high}.
  \item \textbf{Redundancy of wisdom.} Adding wisdom to a detail level helps at \textsf{low},
  but is largely redundant at \textsf{high}.
  \item \textbf{Speed--diversity trade-off.} \textsf{high} sharing accelerates reuse but
  narrows the set of approaches tested, with the net effect depending on problem ruggedness,
  echoing results for human networks \citep{lazer2007network,mason2008propagation}.
  \item \textbf{Same-model homogeneity.} When all roles use one model, organizational
  differences between designs A--E shrink as sharing increases; heterogeneous model assignment
  widens them.
  \item \textbf{Team versus single scientist.} At matched total budget, multi-agent designs
  beat a persistent single scientist only on missions with separable sub-problems, and
  delegation depth beyond one adds coordination overhead faster than useful work.
\end{enumerate}

\paragraph{Protocol.}
I propose a two-stage factorial design. Stage 1 varies lab design (A--F) with sharing held
fixed; Stage 2 varies sharing (the eight configurations) within selected designs, so that
gains from richer sharing are not attributed to governance style. Generation model, request
options, decision mode, task instances, total budget and confirmation protocol are matched,
and design F with no delegation serves as the baseline. Each cell is replicated with fresh
contexts and independent lab runs, since multiple labs sharing a panel are not independent
replications. Outcomes are: verified practical improvement on protected confirmation; total
realized cost including meetings, decisions, children, retries and failures; duplicate
experiments; communication overhead (curation and read cost); the range of mechanisms actually
tested; and whether, after reading a peer view, a prime abandons its approach and whether that
helps. The statistical family and analysis are declared before held-out comparisons are
examined. Labtune's records (configuration identities, read logs of which snapshot entries
entered each context, receipts and costs) are designed to supply exactly these quantities.

\paragraph{Threats to validity.}
Four threats are immediate. First, personas running on one model may differ in name more than
in actual method; the mechanisms they execute must therefore be measured. Second, field-based
sharing filters cannot guarantee semantic isolation: prose admitted into a field may still
encode a recipe, making leakage an outcome to quantify. Third, LLM nondeterminism and
provider-side changes require replication and recorded model identities. Fourth, a synthetic
memory task is a controlled testbed, not evidence that the same organizational effect will hold
in a scientific domain.

\section{Limitations}
\label{sec:limitations}

The largest limitation is also the main boundary of this paper: no empirical comparison of lab
organizations is reported. Labtune currently provides the experimental machinery, not evidence
that any configuration is better. The reference implementation is a bounded phase workflow
rather than a production research service. Live reconfiguration, durable scheduler recovery and
a general asset registry remain planned interfaces. Untrusted candidate execution exists for one
synthetic domain and only under a macOS sandbox; unsupported platforms fail closed. Reviewed
priors currently cover Beta priors for binomial rates in that adapter. Provenance checks can
detect declared overlap but not hidden dependence. Finally, the six lab designs are inspired by
organizational literature rather than validated reproductions of human institutions. A persona
label such as ``experimentalist'' assigns a reasoning role; it is not evidence that the model
possesses human expertise.

\section{Conclusion}
\label{sec:conclusion}

LLM-agent research already has an organization, even when the organization is hidden inside the
software. Labtune makes that choice explicit. A user selects the lab design, controls what can
cross team and lab boundaries, conserves budget under delegation, and separates development
from protected confirmation. Configuration identities, exposure histories and evidence receipts
then preserve enough of the process to ask a more useful question than whether ``multi-agent''
is better than ``single-agent'': \emph{which organization, under which communication policy, for
which kind of problem?}

The broader proposal is to treat artificial research organization as something that can be
studied rather than merely engineered. Human science has long dealt with the tension between
independent search and integration. LLM agents make it possible to reproduce parts of that
tension under controlled conditions. Whether the same organizational regularities appear is an
empirical question. Labtune is intended to provide the apparatus for asking it.

\section*{Code availability}
Source code, lab designs, templates and diagrams are available (version 0.5.0) at
\url{https://github.com/JeranRolevs/labtune}. The software is distributed under the
Labtune Academic Research License 2.0; see the repository for terms.

\section*{Disclosure}
During the preparation of this manuscript, the author utilized a combination of LLM models to rephrase, improve the clarity, readability, and stylistic tone of the introductory draft. After using this tool, the author reviewed, edited, and verified the content to ensure accuracy, and take full responsibility for the final manuscript.

\bibliographystyle{plainnat}
\bibliography{references}

\end{document}